\documentclass[
preprint,
 amsmath,amssymb,
 aps,
]{revtex4-2}

\usepackage{graphicx}
\usepackage{dcolumn}
\usepackage{bm}
\usepackage{subcaption}
\usepackage{ragged2e}
\DeclareCaptionJustification{justified}{\justifying}
\usepackage{amsmath}
\DeclareMathOperator\erfc{erfc}
\begin{document}


\title{Fracture process of composite material in a spring network model}
\author{Haruka Noguchi and Satoshi Yukawa}
\affiliation{%
 Department of Earth and Space Science, Graduate School of Science, Osaka University, Toyonaka, Osaka 560-0043, Japan
}%

\date{\today}
\begin{abstract}
We analyze a two-dimensional spring network model comprising breakable and unbreakable springs. Computer simulations showed this system to exhibit intermittent stress drops in a larger strain regime, and these stress drops resulted in ductile-like behavior. 
The scaling analysis reveals that the avalanche size distribution demonstrates a cut-off, depending on its internal structure. 
This study also investigates the relationship between cluster growth and stress drop, and we show that the amount of stress drop increases in terms of power law, corresponding to crack growth. 
The crack length distribution also demonstrates a cut-off depending on its internal structure. 
The results show that both the cluster growth-stress drop relationship and the crack size distribution are scaled by the quantity related to the internal structure, and the relevance of the exponent that scales the cluster growth-stress drop relationship to the exponent that scales crack size distribution is verified.
\end{abstract}

\maketitle

\section{\label{sec:level1}Introduction} 
Biological materials such as nacre or wood are composed of several materials and constitute a specific structure. It is known that their internal structure serves to improve their mechanical properties, especially fracture toughness. For example, the nacre is a composite material with organic parts and mineral parts arranged in a layered manner; It shows better fracture strength because of this laminated structure \cite{currey1977mechanical, wang2011deformation,nacre2011}. The trial to learn from the structure of biomaterial and exploit it to design a product is called ``biomimicry'' \cite{LURIELUKE20141494}, and it is still an important field of engineering.

In material science, the effects of composition and structure on fracture behavior have attracted attention and have been used to create superior materials for failure. A typical man-made structural composite is fiber-reinforced ceramics, a composite of ceramics and fiber. Even though each component is brittle, fiber-reinforced ceramics show ductile-like behavior \cite{fiber1, fiber2} because fibers prevent crack propagation when cracks propagate in the ceramic phase and meet the fiber. More recently, advancing 3D printing technology has made it possible to create more complex internal structures of composite materials quickly, cheaply, and at a large scale \cite{liu_additive_2021}, and this advancement has gathered much attention from material science to the relationship between structure and failure \cite{al-ketan_mechanical_2017,dimas_tough_2013}. 
For example, Li \textit{et al.} \cite{li_enhanced_2018} created a composite material with a structure consisting of glassy polymer skeletons filled with a highly rubbery thermoplastic elastomer using a 3D printer. By observing the differences in fracture behavior when altering the skeletal structure, they demonstrated that even in composite materials composed of the same type of material, fracture behavior, such as process zone formation, can be controlled by variations in the skeletal structure.

The influence of internal structure can appear not only in fracture behavior but also in scaling behavior. In the layered structures of soft and hard components like nacre, Okumura \textit{et al.} predicted theoretically \cite{okumura_why_2001} and numerically \cite{aoyanagi_stress_2009, hamamoto_realistic_2013} that scaling law with the length of the period between soft and hard layers is valid for the crack tip stress and the crack shape. The other example is about hierarchical structure. Shi \textit{et al.} \cite{shi_scaling_2021} derived the scaling law of yield strength between different hierarchy levels and explained the difference in mechanical properties of the nano-scale hierarchical material in the degree of dealloying. Such ``structure-based scaling relation'' can be a guide to creating composite materials with more complex structures, but it is still not enough for our understanding of how structural properties like the length scale that characterize internal structure appear in fracture behavior.

To bridge this gap, we study the fracture behavior of composite materials, 
especially scaling behavior for the characteristic internal length scale,
with a simple stochastic fracture model by a numerical simulation.
There are several types of stochastic fracture models, such as the fiber bundle model \cite{doi:10.1080/19447027.1926.10599953,pradhan2010failure,hansenFiberBundleModel2015}, the random fuse model \cite{Arcangelis1985ARF, PhysRevLett.110.185505}, the spring-network (SN) model \cite{FBM_1,zapperi2005, PhysRevB.47.695,takesue2010} and so on. 
They are used to understand disorder-induced statistical aspects of fracture 
like the power law of released-energy statistics \cite{garcimartin1997statistical, PhysRevLett.89.185503, PhysRevLett.96.045501, L.I.Salminen_2006}, the self-affine nature of crack morphology \cite{santucci2010fracture,https://doi.org/10.1111/j.1151-2916.1989.tb05954.x,PhysRevLett.78.1062}, intermittent dynamics \cite{PhysRevLett.96.045501,PhysRevLett.101.045501}, and pattern formation \cite{A.Groisman_1994,PhysRevE.60.6449}. 
The base of our model is the spring-network model. The model comprises two spring types: One spring breaks with the application of a specific amount of load and the other is unbreakable under any load, and these two kinds of springs comprise internal structure. As the previous study about the failure of composite material with the stochastic model, Kun \textit{et al.} analyzed the fracture of a random mixture of weak and strong fiber composites by using equal-load sharing and local-load sharing fiber bundle models \cite{hidalgo2008universality, PhysRevE.87.042816}. Tauber \textit{et al.} \cite{PhysRevMaterials.4.063603} considered a spring model that mimics polymer composites. Urabe \textit{et al.} \cite{takesue2010} and Rajesh \textit{et al.} \cite{mayya_role_2018,senapati_role_2023} considered bi-material composite by spring model with two kinds of springs that have different young modulus.
Compared to their models, we used strong springs in our SN model to form a regular matrix structure.

The contributions of this study in terms of determining the effect of the internal structure of composite materials are threefold. First, the present system demonstrates ductile fractures because of its internal structure. Second, the burst size distribution of the present model shows power-law behavior in the intermediate size scale, and it shows a cut-off in the case of the larger avalanche. The scaling analysis showed that the burst size distribution can be scaled according to the size scale determined by its internal structure. Finally, the crack size distribution is scaled by the internal-structure-based crack length, and the stress drop caused by crack growth is scaled by the crack-opening length depending on the internal structure of the material. Our model boasts simplicity in capturing the fundamental properties of the fracture process in the composite material with a matrix structure. Moreover, the model can be used as a prototype for composite materials with a more complex internal structure.

The organization of this paper is as follows. In Sec.~\ref{sec2}, the details of the model and simulation method are presented. Results are described in Sec.~\ref{sec3}, and conclusions are summarized in Sec.~\ref{sec4}.

\section{Model and simulation}
\label{sec2}

\begin{figure}[ht]
\includegraphics[width=60mm]{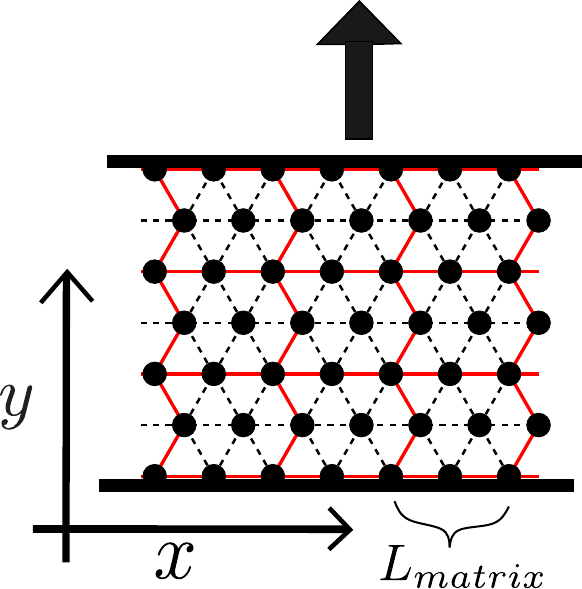}
\caption{Schematic of the spring network model. $N\times N$ particles are located on a triangular lattice, and nearest-neighbor particles are connected using springs. We take the $x$-axis as parallel to the edge of the triangular lattice. The dotted bonds correspond to breakable springs and the red bond corresponds to unbreakable springs. Unbreakable springs are regularly deployed spatially to constitute a $L_{matrix} \times L_{matrix}$ almost square frame. This figure shows the $L_{matrix}= 2$ case. A periodic boundary condition is imposed in the $x$ direction, and a fixed boundary condition is imposed in the $y$ direction for tensile loading. 
\label{fig:model}}
\end{figure}
In this study, we analyzed a two-dimensional SN model, which represents the composite material as 
a network of particles connected by Hookean springs. 
The system comprises $N\times N$ particles on a triangular lattice. 
We take the $x$-axis as parallel to the edge of the triangular lattice. 
All particles possess the same mass, which is taken to be a mass unit, and each pair of nearest-neighbor particles is connected by the Hookean spring, the natural length of which is represented by lattice spacing $l_0$, which is the unit of length. The periodic boundary condition was imposed in the $x$ direction and the fixed boundary condition was imposed in the $y$ direction for tensile loading.
Each spring has a fracture threshold of $l^{*}$, which was randomly selected from a uniform distribution between 0 and 1.
When the strain of the spring, i.e., $\left\lvert l-l_{0}\right\rvert \slash l_0$, becomes larger than the threshold, $l^{*}$, the spring breaks, and it is removed from the system.
In this system, the strains caused by the broken springs were distributed among the remaining live springs to reach the mechanical equilibrium.
The successive breaking of many springs is possible, a phenomenon, which hereafter, is referred to as burst or avalanche. 
The potential energy of this model can be formulated as \cite{takesue2010}
\begin{eqnarray}
    V=\frac{k}{2}\sum_{\langle i,j \rangle}(|\bm{r_{i}}-\bm{r_{j}}|-l_{0})^2g_{ij},
\end{eqnarray}
where $\bm{r_{i}}$ is the position vector of the $i$-th particle, $g_{ij}=1$ indicates a live bond, and $g_{ij}=0$ indicates a broken bond. 
The summing pair $\langle i,j \rangle$ runs the nearest-neighbor pairs on the triangular lattice.
Parameter $k$ is a spring constant, taken as unity.

We replaced some springs with unbreakable springs, $l^{*} = \infty$, for modeling the composite material. 
These unbreakable springs were regularly deployed spatially to constitute a $L_{matrix} \times L_{matrix}$ almost square frame. Here, $L_{matrix}$ is the number of unbreakable springs for one side of the almost square matrix as shown in Fig.~\ref{fig:model}.
For every $L_{matrix}$ layer, unbreakable springs are put parallel to the $x$-axis and zigzag for the $y$-direction.
We call this system the ``matrix-mixture system,'' and we term the system without unbreakable bonds as the ``normal system.''
In this study, we take $N=96$ and $L_{matrix}=6,8,12$.

Next, the system was simulated under strain control as follows. The lowest row of particles was fixed, and a small amount of displacement was implemented among the highest row of particles. 
The system was then allowed to relax to a mechanical equilibrium state. 
The equilibrium state was explored using the FIRE algorithm \cite{bitzek2006structural}. 
After the system relaxed to the mechanical equilibrium state, we decided on which bonds to break. 
If a certain spring's strain was over the fracture threshold, that spring was removed from the system. 
After removing the springs, the mechanical equilibrium configuration was analyzed again without moving the top particles.
This loop was continued until the springs stopped breaking in the mechanical equilibrium state. 
When the system reached this state, we repeated the same procedure. 
The simulation was finally stopped when the system completely broke into two pieces or the strain of the system reached $1$.
For one small uniaxial extension step, the strain of the system increased by $0.1\%$, i.e., $\Delta \epsilon =0.00 01$. Here, the statistically independent 1000 configurations were simulated.

\section{Result}
\label{sec3}
\subsection{Mechanical property}

\begin{figure}[ht]
\includegraphics[width=70mm]{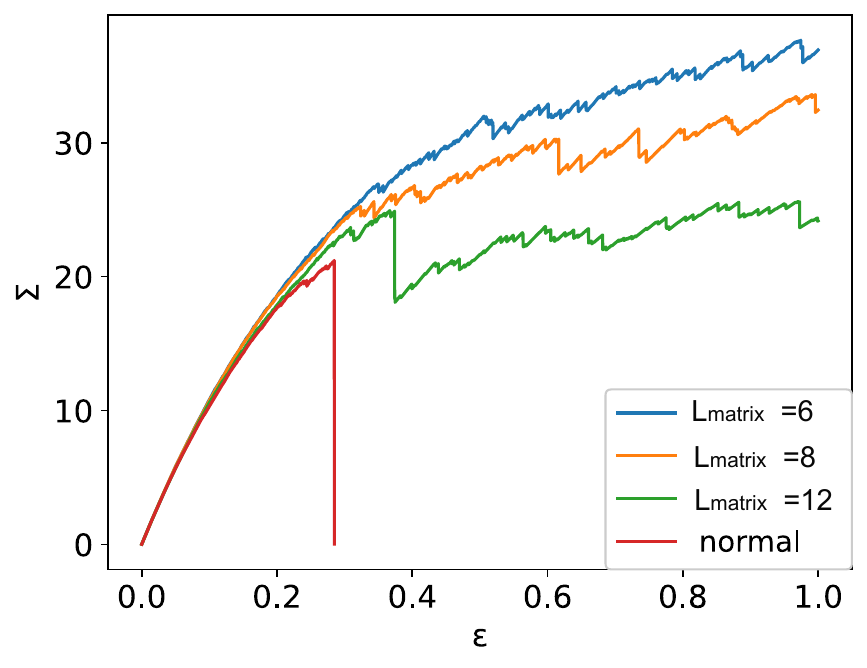}
\caption{Typical stress-strain response of the system. Every system shows a linear response in a small strain regime. Then, the normal system breaks in a brittle manner. The matrix-mixture systems do not demonstrate such an abrupt change but show ductile-like failures with intermittent stress drops. The size distributions of stress drops vary in each system. As the matrix size of the system increases, the larger stress drop also increases, as shown in Fig.~\ref{fig:fig2}.
\label{fig:fig1}}
\end{figure}

\begin{figure}[ht]
\includegraphics[width=70mm]{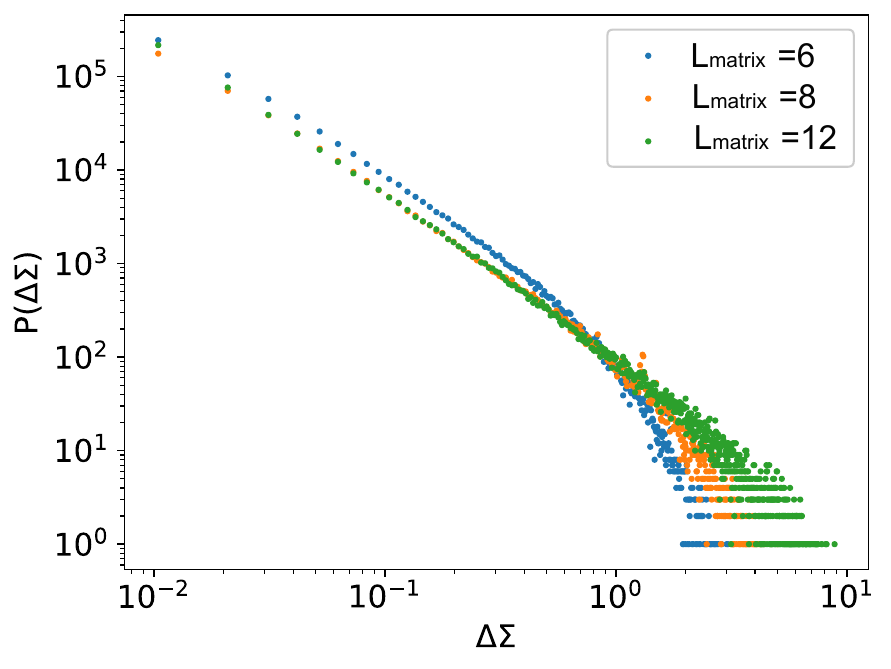}
\caption{Distribution of stress drop, $P(\Delta \Sigma)$. In the smaller $\Delta \Sigma$ regime, $P(\Delta \Sigma)$ shows power-law decay. The larger $\Delta \Sigma$ regime demonstrates a cut-off, depending on $L_{matrix}$. The distribution is considered by the binning with $1/N$.
\label{fig:fig2}}
\end{figure}

We first discuss the stress-strain curve of the system. In this study, we compute
\begin{eqnarray}
    \Sigma=\frac{k}{H_{0}}\sum_{\langle i,j \rangle}(|\bm{r}_{i}-\bm{r}_{j}|-l_{0})\frac{|y_{j}-y_{i}|}{|\bm{r}_{i}-\bm{r}_{j}|} g_{ij},
\end{eqnarray}
as the stress \cite{takesue2010}, where $H_{0}=N l_0$ is the width of the system.
Figure~\ref{fig:fig1} shows the typical stress-strain curve of this system. At the beginning of tension application, all systems show elastic-like behavior, though cracks appear in the system. After that, the slope of the stress-strain curve reduces because of the increase in damage. Eventually, the normal and matrix-mixture systems show completely different mechanical responses. The normal system shows an abrupt stress drop. As shown in previous studies \cite{zapperi1997first,PhysRevE.71.066106}, this stress drop is compatible with crack propagation from one end to another and breaking the system in two. That is, the normal system shows a brittle fracture. 
The matrix-mixture systems show similar behavior at the beginning of loading as the normal system. However, these systems behave as a ductile material in the larger strain regime. 
This ductile behavior could be attributed to the intermittent and instantaneous stress changes. Hereafter, we term this stress change as a stress drop, and its magnitude is denoted as $\Delta \Sigma$.
The ductile regime frequently displays small-scale stress drops, which cancel out the increase in stress.
The comparison of each $L_{matrix}$ shows that the system with a small $L_{matrix}$ value shows less stress drop than the system with a large $L_{matrix}$ value.
To quantify this difference, we investigated the distribution of stress drop, $\Delta \Sigma$, which is denoted as $P(\Delta \Sigma)$, as shown in Fig.~\ref{fig:fig2}. 
All negative instantaneous changes in stress were considered. 
In the smaller stress drop regime, distribution $P(\Delta \Sigma)$ showed power-law decay, and its exponent is almost the same for the different matrix sizes, $L_{matrix}$. In the much larger $\Delta \Sigma$ region, the cut-off was observed to depend on $L_{matrix}$.
The result in Fig.~\ref{fig:fig2} suggests that a smaller matrix significantly suppresses large stress drops.

\subsection{Avalanche}

\begin{figure}[ht]
\includegraphics[width=70mm]{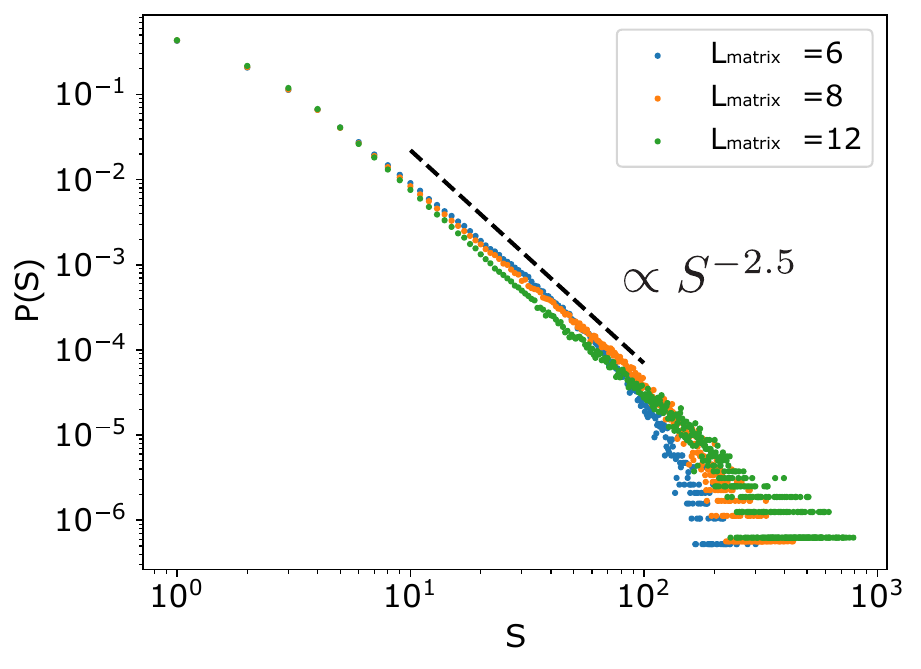}
\caption{Avalanche size distribution. The black dashed line corresponds to the distribution of the normal SN model, $P(S) \sim S^{-2.5}$.
\label{fig:fig3}}
\end{figure}
To understand the effect of suppressing the stress drop by using a matrix structure, we studied the difference between the burst size distribution $P(S)$ between the normal and matrix-mixture SN models, where $S$ is the number of breaking springs during the single loading step with respect to $\Delta \epsilon$. In the normal SN model, the burst size distribution, $P(S)$, behaves as $\sim S^{-\tau}$, and exponent $\tau=2.5$ \cite{zapperi1997first}. 
We plot the burst size distribution in Fig.~\ref{fig:fig3}, wherein all burst events are considered. First, as shown, exponent $\tau$ decreases with the consideration of the internal structure in the matrix-mixture system. 
This indicates that smaller bursts are more likely to occur in the smaller matrix-mixture system.
Second, Figure~\ref{fig:fig3} shows that the burst size distribution demonstrates a cut-off size depending on its matrix size.
Based on these observations, the matrix structure increases the burst events until the intermediate scale and suppresses burst events on a larger scale.

\begin{figure}[ht]
\includegraphics[width=70mm]{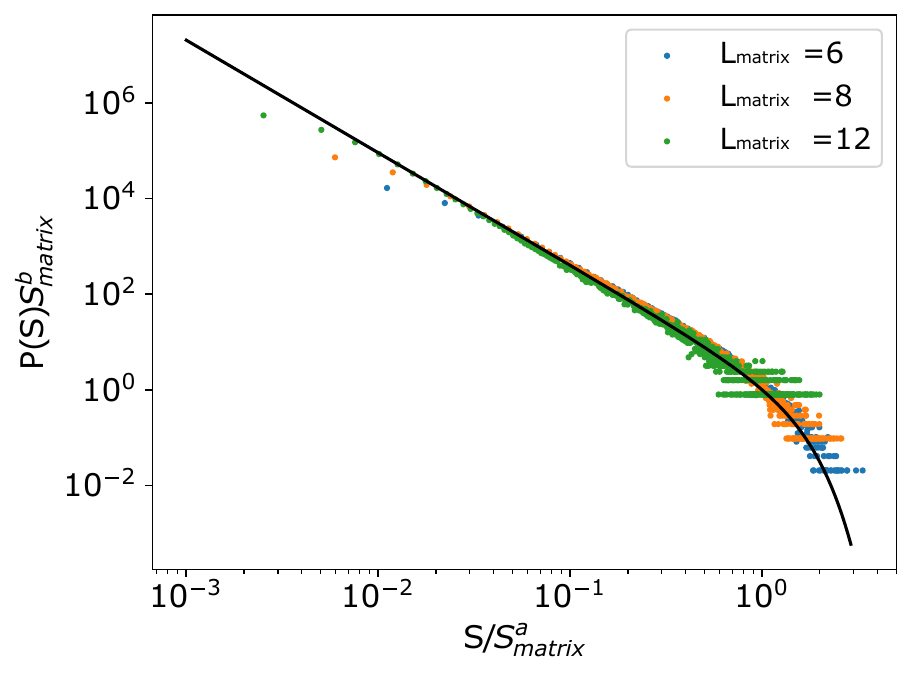}
\caption{Scaled avalanche size distribution. The solid line corresponds to the fitting function, $f(x)=x^{-\tau}\erfc(x-1)$, with respect to the scaling hypothesis shown in Eq.~\eqref{eq:ansatz}. Here, we take scaling variables as $a=1.0$, $b=2.38$, and $\tau=2.35$.
\label{fig:fig4}}
\end{figure}
Next, by using scaling analysis, we discuss the differences among each matrix-mixture system in terms of the burst size distribution.
We assume that the burst size distribution has the following scaling form with exponents $a$ and $b$:
\begin{eqnarray}
P(S,L_{matrix}) = S_{matrix}^{-b} f(S/S^{a}_{matrix}),
    \label{eq:ansatz}
\end{eqnarray}
where $S_{matrix}$ is the number of breakable springs in the matrix calculated as $S_{matrix} = 3 L_{matrix} (L_{matrix}-1)$, and $f(x)$ is the scaling function \cite{PhysRevE.87.042816}.
The result of the scaling analysis showed that the distribution of burst size, $P(S)$, collapses onto the master curve.
We achieved a favorable conformance between the data and master curve $f(x)$, formulated as $f(x) = x^{-\tau}\erfc(x-a)$ with $a=1.0$, $b=2.38$, and $\tau=2.35$. 
This result indicates that the burst event follows the power law with the same exponent, $\tau=2.35$, until the burst size was less than $S_{matrix}$ and showed sigmoidal decay once it increased more than $S_{matrix}$. 
The functional form of $f(x)$ indicates that any apparent change in $\tau$ and the decay behavior in Fig.~\ref{fig:fig4} can be attributed to the difference in $S_{matrix}$. 
These results show that the stress drops were suppressed by the matrix structure because the fracture events were suppressed by the cut-off size, $S_{matrix}$. 

\subsection{Crack coalescence and stress drop}

\begin{figure}
\subcaptionbox{Elastic regime ($\epsilon=0.153,\Delta \Sigma =0.02$)}{\includegraphics[keepaspectratio, scale=0.4]{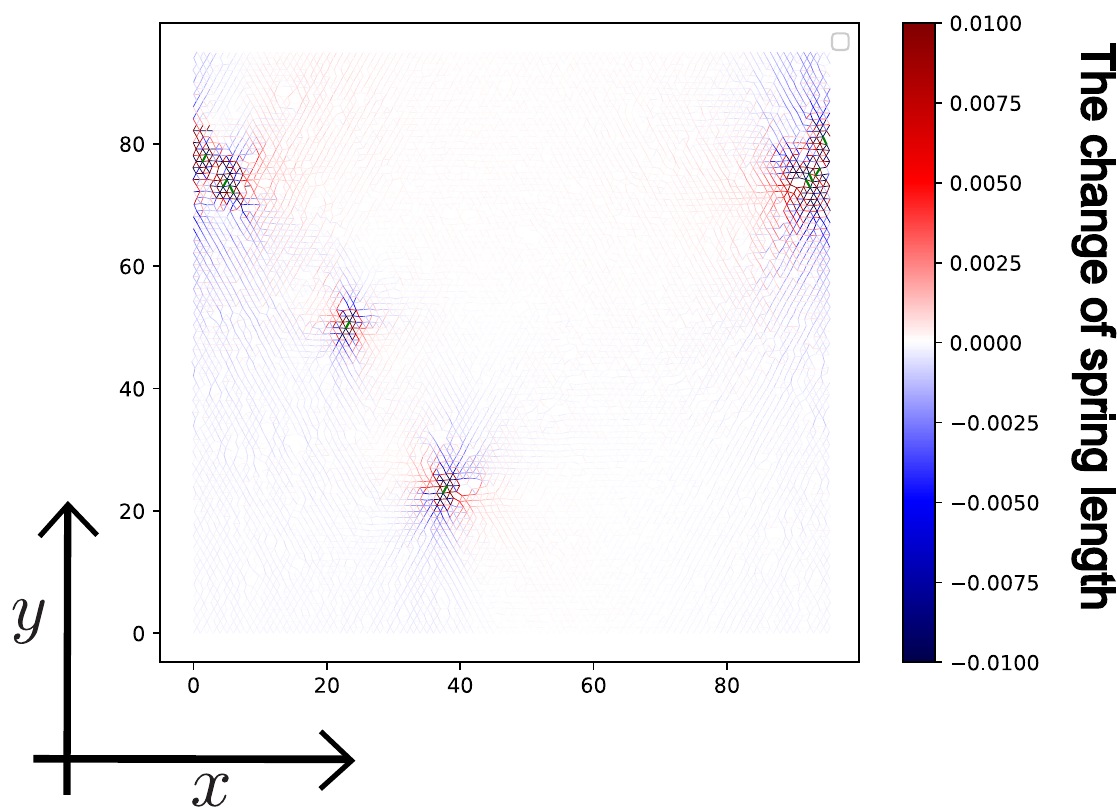}}
\subcaptionbox{Ductile regime ($\epsilon=0.5615,\Delta \Sigma =0.11$)}{\includegraphics[keepaspectratio, scale=0.4]{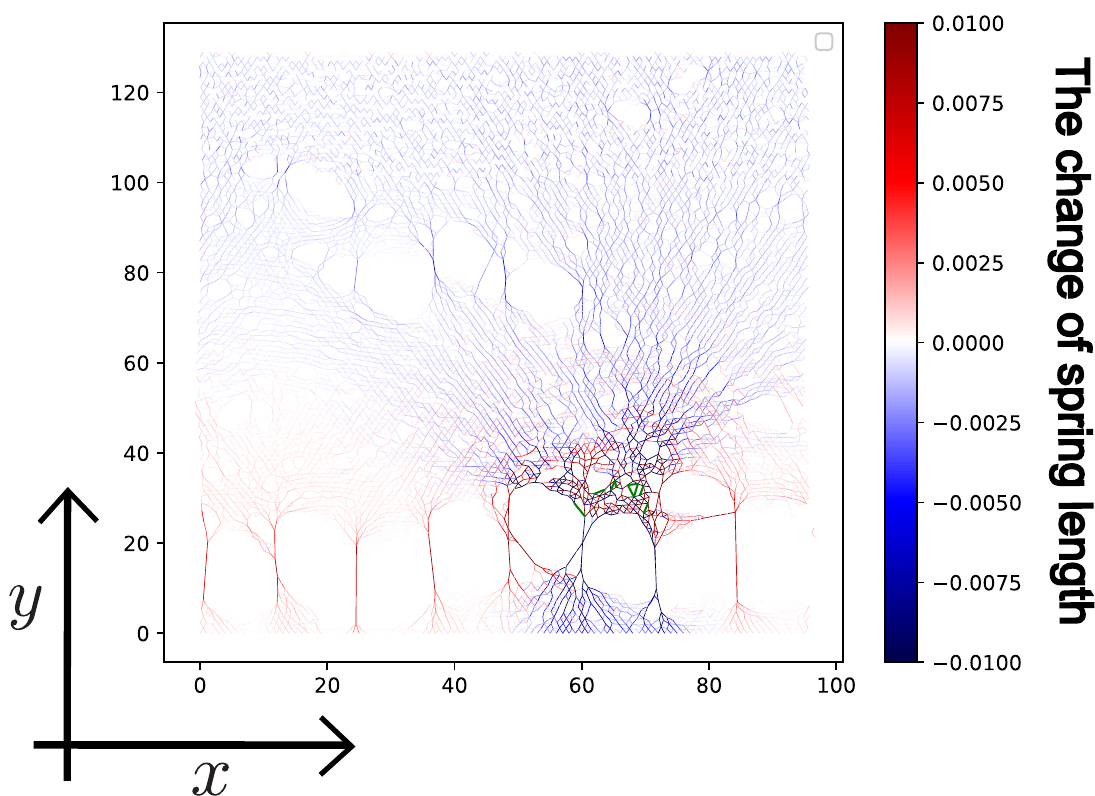}}
   \caption{Snapshots after breaking events in elastic and ductile regimes for $L_{matrix}=12$. 
   Springs with reduced (increased) stress compared to those in the previous step are indicated in blue (red), respectively. 
   Broken springs in the previous step are indicated in green. 
   In both snapshots, the number of breaking springs in the previous step is the same, i.e., 10. 
   The left panel shows a smaller stress drop, which is attributed to the burst events occurring at spatially isolated locations.
   In the right panel, the stress drop is approximately five times larger than in the left panel, and this could be attributed to the fracture events occurring in spatially close locations, accompanied by larger crack growth.\label{fig:fig5}}
\end{figure}
A stress drop depends on not only the number of avalanches but also the spatial distribution of crack formation. 
Figure~\ref{fig:fig5} illustrates the effect of crack coalescence on stress drops.
The figure clearly shows the difference between the spatial distributions of appeared cracks of the elastic and ductile regimes. We could clarify the effects of spatial distribution on the stress drop by observing the internal states of the springs. 
Figure~\ref{fig:fig5} shows the configuration of the SN model at a certain strain rate. The colors represent the changes in the spring length compared with a previous state, i.e., red corresponds to extended springs and blue corresponds to shrunken springs. 
In addition, green represents the springs broken in the previous step.
In both pictures, the number of breaking springs caused by the single tiny displacement is $10$. 
In the elastic regime (Fig.~\ref{fig:fig5}(a)), each appeared crack was spatially isolated. As such, the rupture of springs does not significantly affect the entire structure. 
On the contrary, as the fracturing proceeds to the ductile regime (Fig.~\ref{fig:fig5}(b)), the breaking of springs tends to show a more significant effect on the internal structure by coalescing with existing cracks.

This result suggests that the amount of crack growth is essential for how stress is reduced in certain fracture events \cite{gagnon2001energy}.
Thus, we quantitatively analyzed the relationship between crack growth and stress drop. 

\begin{figure}[ht]
\includegraphics[width=60mm]{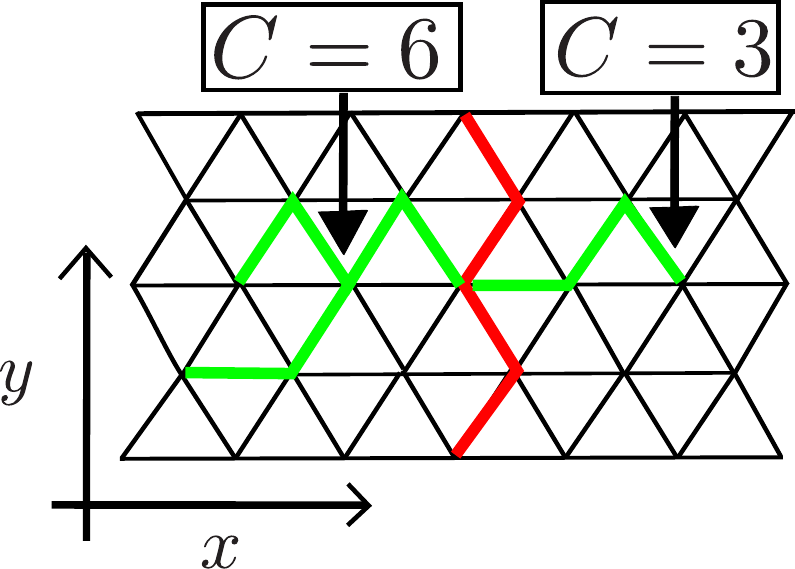}
\caption{Schematic of the definition of the cluster size of cracks. 
The black, green, and red lines correspond to breakable, broken, and unbreakable springs, respectively.
The single crack cluster does not extend over the unbreakable bond. Thus, in this case, there are two clusters: $N_{C=3}=1$ and $N_{C=6}=1$. 
}
\label{fig:figX}
\end{figure}
\begin{figure}[ht]
    \centering
  \includegraphics[width=70mm]{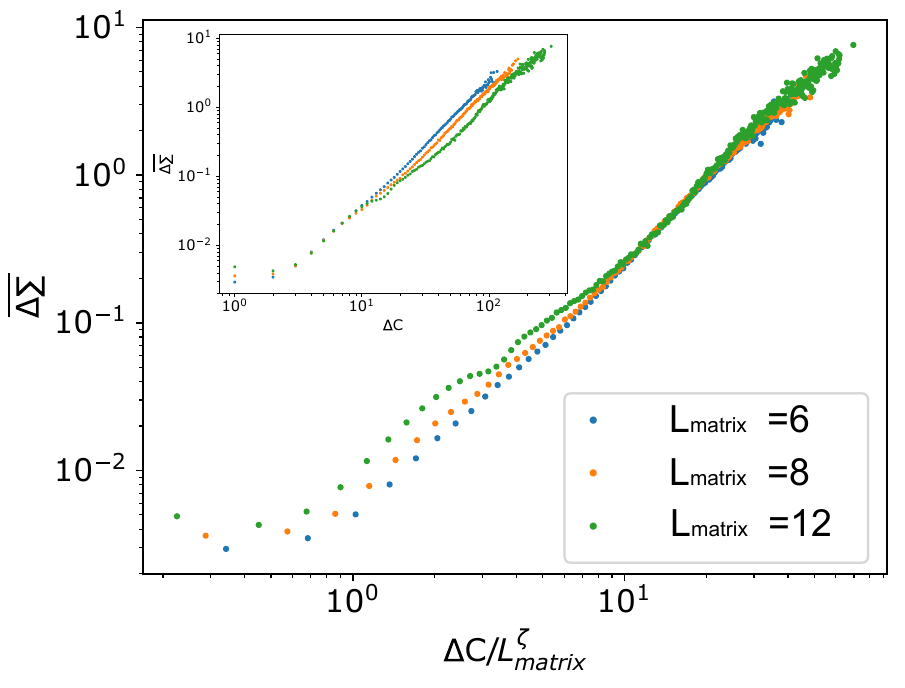}
  \caption{Relationship between the amount of crack growth, $\Delta C$, and the average amount of stress drop for crack growth, $\overline{\Delta \Sigma }$, in terms of matrix size. 
  Inset: Bare relation. All matrix sizes in the smaller $\Delta C \lesssim 10^1$ regime display the same behavior, which deviates in the larger $\Delta C$ regime. 
  Main: Scaled relation. This deviation can be scaled with respect to $\Delta C$ divided by $L_{matrix}^{\zeta}$, where the exponent is $\zeta=0.55$.
  For plotting, the horizontal axis, $\Delta C$, has been binned by an integer to reduce fluctuations.}
  \label{fig:fig6}
\end{figure}
To quantify the effect of crack growth, we introduced two quantities, $C$ and $N_{C}(\epsilon)$, where $C$ corresponds to the size of the crack cluster and $N_{C}(\epsilon)$ is the number of crack clusters with size $C$ at strain $\epsilon$, as shown in Fig.~\ref{fig:figX}.  
In this study, we assumed that the single crack cluster does not extend over the unbreakable bond. If the cluster seems to extend over the unbreakable bond, we have identified it as two separate clusters.

Based on the percolation theory \cite{staufer}, the increment in crack cluster size, $\Delta C $, according to the amount of crack growth is defined as   
\begin{eqnarray}
    \Delta C=\sqrt{\sum_{C}\left( C^2N_{C}(\epsilon)-C^2N_{C}(\epsilon-\Delta \epsilon)\right)}.
\end{eqnarray}
This quantity is the square root of the difference between the average moments of crack cluster size before and after a tiny loading step in a sample. It is uniquely determined for a tiny loading step and one sample. 
The stress drop $\Delta \Sigma$ is also uniquely determined for a tiny loading step.
Thus, the stress drop $\Delta \Sigma$ during a tiny loading step has one-to-one correspondence with $\Delta C$. 
The increment of the crack cluster size shows a large value for the large clusters of broken springs and a small value for several small clusters, even with the same number of broken springs. 
Figure~\ref{fig:fig6} shows the relationship between $\Delta C$, binned by an integer, and the average stress drop $\overline{\Delta\Sigma}$ at each bin over the samples during the whole loading process $\epsilon=0$ to $1$.
Its definition is as follows:
\begin{eqnarray}
    \overline{\Delta \Sigma}=\frac{\sum_{sample}\sum_{\Delta C \in bin} \Delta \Sigma}{\sum_{sample}\sum_{\Delta C\in bin}1}.
\end{eqnarray}
The amount of stress drops roughly increases in a power-law manner.
The inset of Fig.~\ref{fig:fig6} shows the behavior of the stress drop to deviate near $\Delta C \sim 10^1$.
This deviation can be scaled with respect to $\Delta C$ over $L_{matrix}^{\zeta}$, as shown in Fig.~\ref{fig:fig6}. 
Exponent $\zeta \simeq 0.55$ results in the best fit in regime $\Delta C >30$.

\begin{figure}[ht]
    \centering
\includegraphics[width=70mm]{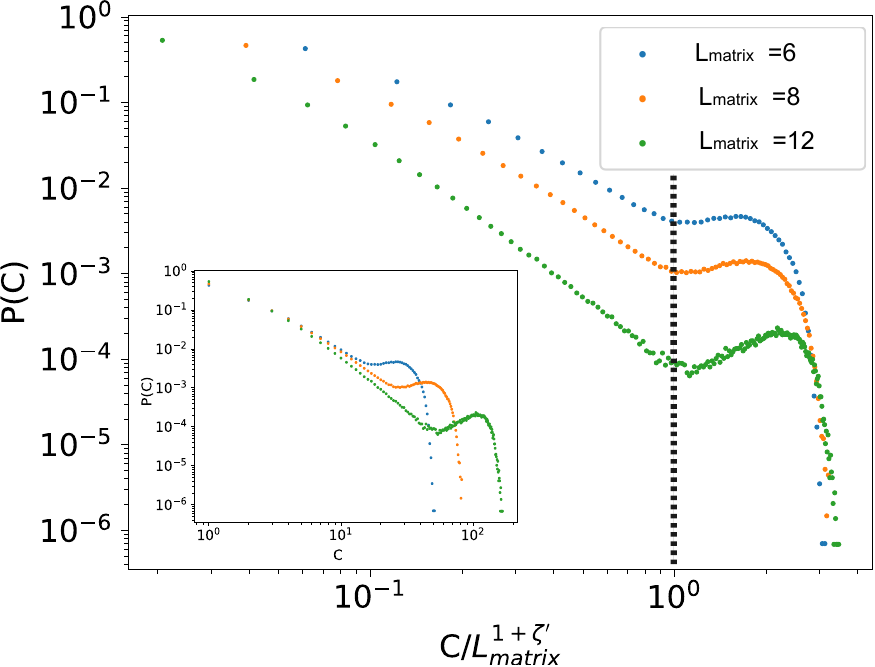}
  \caption{Crack size distribution for each matrix size at the final state, $\epsilon=1$. 
  Inset: Bare distribution. This distribution follows the power law in small-sized cracks. A cut-off size corresponding to its matrix size is observed in the large-sized cracks. 
  Main: Scaled distribution. By scaling the crack size with respect to $L^{1+\zeta^{\prime}}_{matrix}$, we get the point at which deviation from the power law begins and the cut-off size can be scaled independently on the matrix size.
  Exponent $\zeta^{\prime}=0.56$ is almost the same as $\zeta$ in Fig.~\ref{fig:fig6} The black dashed line corresponds to $C/L_{matrix}^{1+\zeta^{\prime}}=1$. 
  \label{fig:fig7}}
\end{figure}
Finally, we analyzed the crack size distribution, $P(C)$, for each matrix at the final state, $\epsilon=1$.  Here $P(C)$ is defined as
\begin{equation}
    P(C)=\frac{\sum_{sample}N_{C}(\epsilon=1)}{\sum_{sample}\sum_{C}N_{C}(\epsilon=1)}
\end{equation}.
In the inset of Fig.~\ref{fig:fig7}, the distribution shows the power-law decay in the small crack length regime. 
This decay is consistent with that observed in previous studies \cite{PhysRevE.65.056105}. 
In this regime, the matrix structure does not affect the crack size. The effect of matrix structure becomes apparent in the region with large cracks. The distribution shows a cut-off corresponding to the matrix size for a long crack. This result clearly shows that the matrix structure of the system suppresses crack growth.
By scaling the crack size by using $L^{1+\zeta^{\prime}}_{matrix}$, the point at which deviation from the power law begins and the cut-off size can be scaled independently on the matrix size is shown in Fig.~\ref{fig:fig7}.
We achieved the best fit by using exponent $\zeta^{\prime}=0.56$,
and this is almost the same as $\zeta$ in Fig.~\ref{fig:fig6}.
We ascribe this agreement between $\zeta$ and $\zeta^{\prime}$ to corresponding $L_{matrix}^{\zeta}$
with an effective crack width. Previous research \cite{PhysRevE.61.6312} showed that the roughness exponent of the random fuse model was $2/3$, and it was $0.62$ for the Born model \cite{guido1998}; These values are close to the value achieved in the current study: $\zeta\ \simeq 0.55$. 
Considering $L_{matrix}^{\zeta^{\prime}}$ as the characteristic length of the crack width in the matrix, $L_{matrix}^{1+\zeta^{\prime}}$ can be identified according to the typical crack length in the matrix. 
The crack length distribution follows a power law similar to that of the normal SN model for small crack lengths. However, owing to the matrix structure, 
the cracks cannot grow larger, and a peak appears in the distribution after the scale of $L_{matrix}^{1+\zeta^{\prime}}$.

\section{Conclusion}
\label{sec4}

In summary of our results, we analyzed the fracture process of composite materials and their statistical properties according to the internal structure of the material by using the SN model with a mixture of breakable and unbreakable springs. 
We found that the proposed SN model shows ductile-like fractures because of intermittent stress drops. In addition, we revealed that avalanche size distribution is well scaled by the number of springs in matrix $L_{matrix}(3L_{matrix}-1)$. The scaling function can be written as $x^{-\tau}\erfc(x-1)$, suggesting that the fracture event follows the power law, similar to the normal SN model \cite{zapperi1997first}, and it decays abruptly when reaching a specific number of springs in a matrix. 
We also revealed the relationship between crack cluster growth and stress drop. Larger clusters appeared when cracks merged, resulting in a more significant stress drop. On average, the amount of stress drop increased based on the power law, followed by the growth in crack clusters. The cluster size distribution showed a cut-off corresponding to the matrix size, and a typical crack length in the matrix could rescale the size cut-off. The matrix size limits the size of the cluster.

This study showed that the material's internal structure, the regular matrix structure, affects the fracture behavior under quasi-static tensile stress. In particular, the length scale of the internal structure significantly controls the fracture behavior by scaling. Interestingly, these properties appear in the present simple model. Additionally, the simplicity of the present model enables experimental verification of the present results using 3D printing techniques \cite{liu_additive_2021}. 
It is intriguing to consider how the matrix length scale appears in the other type of fracture. For example, how does the effect of the length scale of the internal structure appear in the statistical law of fragmentation \cite{PhysRevLett.96.025504, PhysRevE.77.051302} or fatigue failure \cite{PhysRevLett.100.094301}? 
We can investigate these problems with an extension of the present model 
because of its simplicity of the present model.


\begin{acknowledgments}
The authors thank T. Hatano and N. Sakumichi for the fruitful discussion and useful comments.
This work was supported by JSPS KAKENHI Grant Number 19K03652.
\end{acknowledgments}

%

\end{document}